# Modern approaches to improving phase contrast electron microscopy

Jeremy J. Axelrod[1,2], Jessie T. Zhang[1], Petar N. Petrov[1,2], Robert M. Glaeser[3], Holger Müller[1,2]

1 Department of Physics, University of California Berkeley, Berkeley, CA 94720, USA

2 Lawrence Berkeley National Laboratory, One Cyclotron Road, Berkeley, CA 94720, USA

3 Department of Molecular and Cell Biology, University of California Berkeley, Berkeley, CA 94720, USA

Although defocus can be used to generate partial phase contrast in transmission electron microscope images, cryo-electron microscopy (cryo-EM) can be further improved by the development of phase plates which increase contrast by applying a phase shift to the unscattered part of the electron beam. Many approaches have been investigated, including the ponderomotive interaction between light and electrons. We review the recent successes achieved with this method in high-resolution, single-particle cryo-EM. We also review the status of using pulsed or near-field enhanced laser light as alternatives, along with approaches that use scanning transmission electron microscopy (STEM) with a segmented detector rather than a phase plate.

## History of phase plates in transmission electron microscopy

A phase plate generates image contrast for transparent (non-absorbing) objects, which can be achieved only imperfectly by defocusing the image. Phase plates generate contrast by introducing a relative phase shift between the scattered component of the electron beam's wave function and the component that has not been scattered by the sample (see figure 1). This can be achieved by applying a phase shift at the center of a diffraction plane of the transmission electron microscope (TEM), where the unscattered wave is focused to a small spot. A 90° phase shift applied to just that spot maximizes the contrast.

The phase plate was originally invented for optical microscopy by Zernike in the early 1930s [1]. Boersch applied the idea to TEM [2]. Many experimental attempts were made to implement not only two of Boersch's proposals but also a large number of other designs, as is reviewed by [3]. None of these devices, however, made it past the "proof of concept" stage, with the exception of the Volta phase plate [4], which has since been used in multiple investigations, too numerous to be reviewed here.

The Volta phase plate is a heated continuous carbon film, which builds up a patch potential at the point where it is struck by the unscattered beam in the electron diffraction pattern. The film must be heated to ~200 °C to prevent the buildup of charged contaminants. The patch potential produces a localized phase shift, in addition to that produced by the atomic potentials of the continuous carbon film. However, the phase shift continues to rise as the carbon film is irradiated (such as during data taking), thus causing the amount of relative phase shift to increase beyond the desired value of 90°. It is therefore necessary to periodically move the Volta phase plate from one position to another during use, to restart the build-up of patch potentials. As a result, the Volta phase plate is somewhat cumbersome to use; one must first wait for the Volta potential to build up each time it is moved, and its useful lifetime at each new position is limited, as is shown in figure 1 of [5]. In addition, the Volta phase plate also introduces an unwanted contrast transfer function envelope [6], limiting its performance at high resolution. As a result, it is difficult to obtain any performance benefits in single-particle cryo-EM from the Volta phase plate, apart from improving the contrast [7]. As a result, the Volta phase plate now seems to be less frequently used, and there remains considerable scope for developing a more successful way to generate phase contrast in electron microscopes.

## The laser phase plate

The laser phase plate (LPP) applies a 90° phase shift to the unscattered wave using a high-intensity laser beam, which is focused in the diffraction plane. The interaction between the oscillating electromagnetic field of the laser beam and the charge of the electron results in a repulsive effective potential known as the ponderomotive potential [8], [9]. This potential (like any other) imparts a phase shift to a through-going electron wave function. The LPP stands in contrast to previous phase plate designs in that it does not require any material to be placed in or near the electron beam, which has been a root cause of most of the issues which plague these previous designs. However, the ponderomotive potential is relatively weak, and designing a successful LPP requires achieving sufficient laser intensity to provide a 90° phase shift. Several different designs for an LPP have been proposed.

### Pulsed

A pulsed laser can generate the required intensity during each pulse (typically, <1 ns). However, the electron source must also be pulsed and synchronized such that the electron pulses overlap in space and time with the laser pulses. Pulsed electron sources typically rely on photoemission using an ultraviolet laser pulse [10], [11], though continuous sources can be chopped by deflecting the beam using radio-frequency (RF) fields [12], [13].

Two designs for a pulsed laser phase plate (P-LPP) have been proposed. In the first, the laser beam axis lies in the diffraction plane, and the laser and electron beams intersect perpendicularly, with the laser pulses focused at the center of the diffraction plane (see figure 2a). van Leeuwen et al. have calculated the phase shift imparted to the electron beam using this configuration [14]. They find that a 90° phase shift is readily achievable using current pulsed-laser technology (e.g. ~10 nJ, 100 fs pulses with a central wavelength of 800 nm). However, if the laser beam is focused to a width of less than roughly one optical wavelength and the electron beam kinetic energy is greater than roughly 200 keV, the phase shift decreases and the electron beam starts to strongly inelastically scatter from the laser beam. Inelastic scattering would degrade the imaging properties of the phase plate since the unscattered wave would not coherently interfere with the scattered wave to form a phase contrast image. Using a slightly larger focus (e.g. 1.25 times the wavelength) and setting the laser polarization perpendicular to the electron beam axis avoids this problem. Du and Fitzpatrick further consider the case of two perpendicularly intersecting laser beams, which somewhat reduces the effective size of the phase shift spot, especially if the laser beam focal waists are substantially larger than one wavelength [15]. They also conclude that the necessary laser intensity is achievable with current technology.

The second P-LPP design is one in which the axes of the electron and laser beams are nearly aligned. The laser pulses co- or counter-propagate along the electron beam axis, and are focused in the diffraction plane where they temporally overlap with the electron pulses. This configuration is likely more difficult to implement in a TEM because mirrors must be placed near the electron beam to achieve co-/counter-propagation. Figure 2b illustrates the case of a counterpropagating beam, where the mirrors each have a hole to pass the electron beam. The co-/counter-propagating configuration has the advantage that the shape of the phase shift applied in the diffraction plane can be patterned by controlling the shape of the laser beam, for example using a spatial light modulator. de Abajo and Konečná proposed that this could be used as an aberration corrector or arbitrary electron beam shaper [16]. Mihaila et al. then experimentally demonstrated a P-LPP as an adjustable focal length (converging or diverging) electron lens, and as a (nearly) arbitrary beam shaper [17]. They did not demonstrate phase contrast imaging, as their experiment used a modified scanning electron microscope that was incapable of recording phase contrast images of a sample.

The primary challenge facing all P-LPP designs is achieving a sufficiently high electron source brightness. Femtosecond pulsed electron sources, whether using photoemission or RF chopping, can only operate with an average of ~1 electron per pulse, since Coulomb repulsion between multiple electrons within a pulse causes spatiotemporal decoherence of the beam, resulting in loss of image resolution [18], [19], [20]. Longer pulses reduce the effect, but increase the P-LPP's pulse energy requirement to achieve a 90° phase shift [11]. As such, pulsed source TEMs have so far only demonstrated resolutions of ~1 nm [21] compared with 0.2 nm or better when using a continuous source. The source brightness is therefore effectively limited by the pulse repetition rate of the P-LPP. State-of-the-art pulsed sources have a brightness of ~$3.6 \times 10^{20}$ electrons/sr/m$^2$/pulse [21], [22], [23] compared to continuous sources (Thermo Fisher Scientific X-FEG) at ~$1.9 \times 10^{32}$ electrons/sr/m$^2$/second. Therefore, unless the brightness of pulsed sources can be improved, achieving similar coherence and flux as a continuous source would require an extremely (perhaps impossibly) high pulse repetition rate of $5.3 \times 10^{11}$ Hz. Accepting a lower source brightness may ultimately restrict P-LPPs to applications which are amenable to ~1 nm resolutions or long exposure times (minutes rather than seconds).

## Near-field

Near-field laser phase plates use the interaction between electrons and photons close to a surface or nanostructure. In the evanescent waves generated in the near-field, single photons can transfer momentum between the surface and electrons, thereby phase shifting the electrons using a much lower optical intensity. This concept was originally used to image light fields using electron beams

(photon-induced near-field electron microscopy - PINEM) [24], [25], and later extended as a means to actively control electron beams [26].

There has been recent experimental work demonstrating the necessary components to create such a phase plate. In one approach, a pulsed electron beam passes through a surface with a laser pulse reflecting off it. The evanescent waves can be shaped either by the grating structure of the material [27], or by shaping the incident light field using a spatial light modulator [28]. In the latter case, a Hilbert phase plate is theoretically proposed by patterning half the interaction area to generate a 180° phase shift. In another approach, the electron beam passes in close proximity to a microresonator that supports whispering gallery modes that can interact with the electrons [29]. The advantage of this approach is that it uses a continuous-wave (CW) laser which can phase shift a continuous electron beam. However, all near-field experiments to date demonstrate substantial inelastic scattering of the electrons, an unwanted feature for a phase plate. Additionally, by the nature of the near-field interactions, these approaches all require the electron beam to either pass through or near a surface. This can lead to charging of the material and unstable imaging properties, so it remains to be seen whether this approach will be feasible for phase contrast microscopy of biological specimens.

## Continuous-wave

The continuous-wave laser phase plate (CW-LPP) is the only type of LPP that has so far been demonstrated. It uses an optical cavity consisting of two highly reflective (~99.99%) mirrors to resonantly enhance the intensity of a CW laser beam to the value needed to generate a 90° phase shift (see figure 2c). This approach requires state-of-the-art laser optical components, and the current version took roughly 10 years to develop [30], [31], [32], [33], [34]. Since the laser beam is continuously present, the CW-LPP can be used with either pulsed or (the much more common) continuous-current TEM electron sources.

The current generation of the CW-LPP can achieve continuous operation and stable phase shifts in experiments that extend over many hours, with minimal human intervention (see figure 3). In typical operation, a 300 keV electron beam can be phase shifted by 90° with 10 W input laser power at 1064 nm wavelength, resonantly enhanced to 75 kW within the cavity. An example of images of an apoferritin single-particle analysis cryo-EM sample taken with and without the CW-LPP is shown in figure 4. This CW-LPP is installed in a Thermo Fisher Scientific Titan TEM which has been custom-designed with an additional column section which provides the physical space for the CW-LPP. Additional electron relay optics provide a magnified diffraction plane at the location of the CW-LPP, which reduces the cut-on frequency of the phase plate by a factor of 5.7. In this configuration, the phase plate provides a cut-on spatial frequency of ~0.004 $nm^{-1}$ (defined as the spatial frequency where the magnitude of the azimuthally-averaged contrast transfer function first exceeds 0.5). An earlier, resolution-limiting problem with this design, caused by thermal magnetic field noise from the electron beam liner tube of the CW-LPP, was solved by increasing the diameter of the electron beam liner tube in the phase plate [34]. The relay optics also introduce additional spherical and chromatic aberrations that are best compensated for by using a spherical aberration corrector and a cold field-emission gun.

As a side effect, the standing wave of the laser beam in a CW-LPP diffracts the unscattered electron beam and forms ghost images that are offset by a fixed distance (see figure 4 of [31]). The ghost images are relatively weak for thin samples but it remains to be determined whether they pose a problem for thicker samples. Ghost images can be eliminated, if needed, by using a particular polarization of the laser beam of the CW-LPP at the expense of higher cut-on frequency [9]. The higher cut-on frequency could in principle be offset by using a cavity with a smaller focal waist, but this requires further development of the optical cavity technology. Another approach that is being pursued is to use two cavities crossed perpendicular to each other, which increases the number of ghost images but decreases their intensity because each laser beam only needs to provide a 45° phase shift.

The CW-LPP, with its stable operation, promises to be the primary choice of LPP. A spherical aberration-corrected electron microscope with a CW-LPP is soon to be installed at the University of California, Berkeley, and efforts to commercialize the CW-LPP are also underway. In addition to the crossed cavities and cavities with smaller focal waist mentioned previously, miniature versions of the CW-LPP that can be retro-fitted to existing electron microscopes without the need for relay optics are also being developed, which would expand the usage of the CW-LPP in the cryo-EM community.

# Alternatives to using phase plates

## Scanning-transmission electron microscopy (STEM)

While scanning-transmission electron microscopes are most commonly used for dark-field imaging and for electron energy-loss spectroscopy, they can also be used to produce images from which one computationally recovers the phase of the exit wave. Ptychography, originally conceived as a way to solve the "phase problem" in X-ray crystallography [35], is readily implemented in a scanning transmission electron microscope (STEM) that is equipped with a pixel-array area detector [36], [37]. An alternative approach is to use a segmented, bright-field detector [38] to produce images that approach what can be achieved by conventional TEM with a phase plate, depending upon how finely segmented the camera is.

## Ptychography

In electron ptychography, the probe beam is focused such that it covers a small area on the sample, and electron diffraction patterns are recorded from partially overlapping areas as the beam is scanned across the sample. Since the successive diffraction patterns are not independent of one another, i.e. they are derived from irradiated areas that are partly but not entirely the same, recovery of phases as well as amplitudes from the measured diffraction intensities is mathematically well-posed.

Low-dose ptychography has only recently been applied to radiation-sensitive specimens [39], [40], and addressing its potential limitations is still in its early days. For example, the issue of there being a limited bandwidth over which information is recovered efficiently was initially described in figure 2 of [41]. To overcome this limitation, it was subsequently proposed that data be merged for images of macromolecular particles recorded with different convergence angles, similar to merging data recorded with different defocus values in conventional cryo-EM. When this approach was first tried experimentally, however, the resolution was less than expected from simulations, a result that the authors attributed to beam-induced motion [42].

## Segmented-detector bright-field STEM

Information about the structure of a sample is also contained in the bright-field region of a STEM detector, where interference between the unscattered and scattered electron beams occurs when the diameter of the focused probe is smaller than roughly half the reciprocal of the spatial frequency of interest. Within the region of overlap, constructive and destructive interference then causes the intensity to oscillate periodically as the position of the focused probe is scanned relative to the phase origin of the spatial frequency of interest. The use of a detector split into independent half planes can avoid the situation that the oscillating signals produced by a scattered beam and its Friedel mate cancel one another. Further dividing the half planes into separate annular and radial segments can partially avoid the incoherent summation of (oscillating) intensities generated for different spatial frequencies, depending upon how small the segments are; for further details see [43]. From this explanation, it is clear that increasing the number of segments also increases the ability to recover phase-contrast information. In the limit, a highly pixelated camera produces a 2D set of measurements for each pixel in the 2D image of the specimen, leading to what is called 4D-STEM. The downside of 4D-STEM is that it places high demands on the speed of data readout as well as the cost of data processing and storage.

At present, 16-sector STEM detectors are offered commercially (the Panther detector from Thermo Fisher Scientific, and the Opal detector from elmul), and highly pixelated cameras are available from a number of vendors. The independent signals from different sectors can be merged in many different ways, one of which is referred to as producing "integrated differential phase contrast" (iDPC). The iDPC method shares the desirable feature of phase-contrast conventional TEM that image intensities of weak-phase objects are linear in the projected potential.

First results using a 4-sector detector have recently been obtained with biological samples [44]. While these results were not as good as those obtained by conventional cryo-EM, one can expect to see significant improvement as the number of sectors is increased and as methods are added to overcome beam-induced motion. First results for beam-sensitive materials have also been obtained using 4D-STEM [45], but this has not yet been attempted with biological samples.

# Acknowledgments

This work was supported by the U.S. National Institutes of Health (grant no. 2R01GM126011), Chan Zuckerberg Initiative through the Silicon Valley Community Foundation, Gordon and Betty Moore Foundation (grant no. 9366), and a cooperative research and development agreement (CRADA) with Thermo Fisher Scientific (award number AWD00004352). Grant no. 2R01GM126011 and award number AWD00004352 were administered at Lawrence Berkeley National Laboratory under Contract No. DE-AC02-05CH11231. P.N.P. acknowledges support from a postdoctoral fellowship from the National Institute of General Medical Sciences of the National Institutes of Health under award number F32GM149186. The content is solely the responsibility of the authors and does not necessarily represent the official views of the National Institutes of Health.

## Of special interest

[23] Types of pulsed photoemission electron sources are reviewed. Different methods for manipulating the electron beam using lasers are discussed, including direct laser acceleration and temporal phase modulation.

[28] Demonstrates the use of a near-field pulsed laser phase plate for electron beam shaping. A laser beam is shaped using a spatial light modulator and illuminates a thin gold and silicon nitride film through which the electron beam passes. The inelastically scattered electrons are imaged using an energy filter.

[29] Temporal phase modulation (inelastic scattering) of a continuous electron beam is demonstrated using the continuous-wave evanescent field of a silicon nitride ring microresonator. Only microwatts of laser power are required to inelastically scatter most of the electrons due to the strong coupling between the evanescent field and electron beam.

[44] Cryo-electron microscopy reconstructions of keyhole limpet hemocyanin and tobacco mosaic virus are determined to resolutions of 6.5 Å and 3.5 Å, respectively, using integrated differential phase contrast scanning transmission electron microscopy (iDPC-STEM).

## Of outstanding interest

[14] A pulsed laser phase plate is proposed which uses femtosecond laser pulses and a pulsed electron source generated by radio-frequency chopping and compressing of a continuous (field emission) electron source. The validity of the ponderomotive potential approximation is investigated theoretically and shown to hold so long as the laser beam focal waist is larger than roughly one laser wavelength. The authors conclude that commercially available off-the-shelf pulsed lasers are capable of generating a 90° phase shift.

[15] A pulsed laser phase plate is proposed which uses picosecond laser pulses and a photoemission-driven pulsed electron source. The case of two perpendicularly intersecting laser beams is also considered. The authors show that while commercially available off-the-shelf pulsed lasers are capable of generating a 90° phase shift when a synchronized pulsed electron source is used, such a system cannot provide the necessary phase shift to all electrons in a continuous electron source.

[17] A counterpropagating femtosecond pulsed laser phase plate is experimentally demonstrated using a pulsed photoemission electron source in a scanning electron microscope. The phase plate uses a spatial light modulator to generate programmable phase shift profiles. Phase shift profiles generating converging/diverging lenses and higher order aberrations are demonstrated, as is nearly arbitrary beam shaping. The authors do not directly demonstrate phase contrast imaging due to limitations of their electron optics.

[34] Thermal magnetic field noise emanating from the conductive materials used in the mount for a continuous-wave laser phase plate (CW-LPP) is experimentally shown to cause resolution loss which limited the performance of previous CW-LPP prototypes. The resolution loss is largely removed by using a mount designed with a larger diameter electron beam liner tube (8 mm versus 2 mm previously).

[40] Ptychographic reconstruction of frozen hydrated apoferritin to a resolution of 5.8 A.

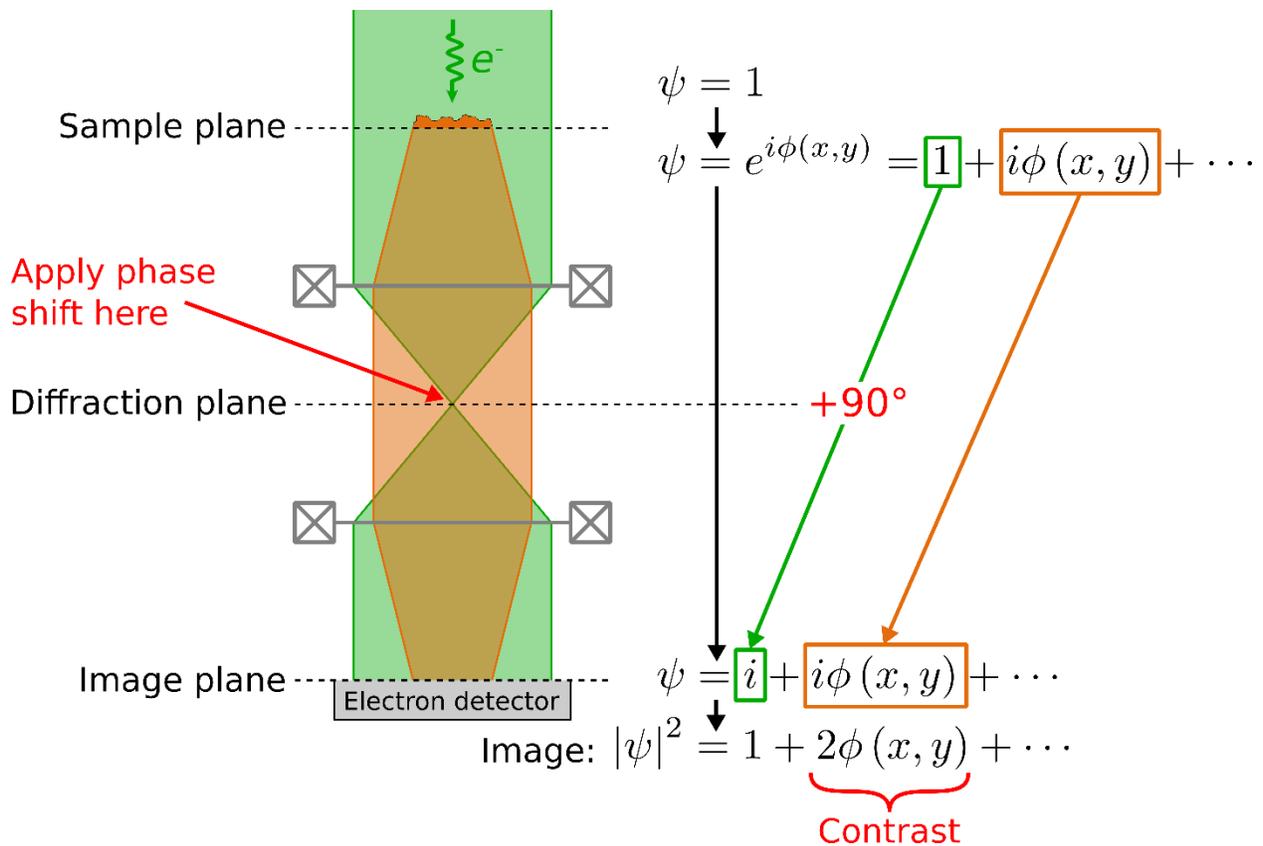

**Figure 1: The principle of phase-contrast TEM imaging.** A planar electron wave function (denoted by $\psi = 1$) is incident on a sample. Propagation in the +z-direction through the sample spatially phase-modulates the electron wave by phase $\phi(x, y)$, while having a negligible effect on its amplitude. A phase plate then applies a 90° phase shift to the unscattered component of the wave (green) relative to the unscattered wave (orange), which causes them to interfere in the image plane, generating an image with phase contrast.

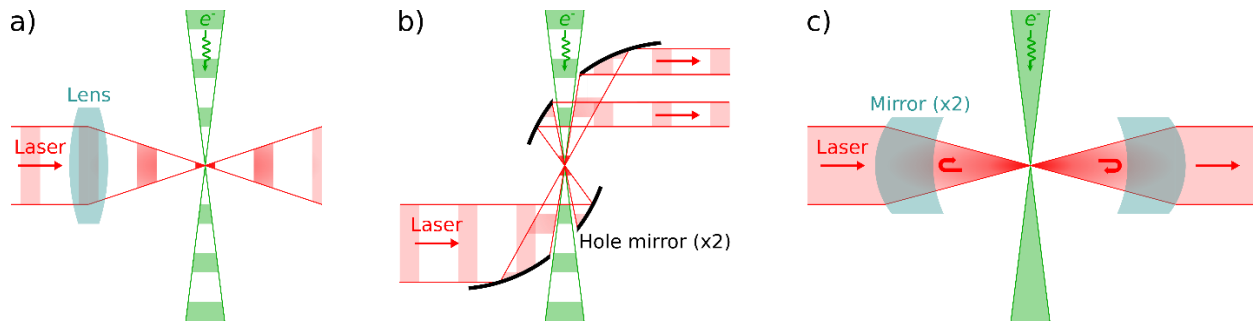

**Figure 2: Continuous-wave and pulsed laser phase plate designs.** a) Pulsed design where the laser and electron beams intersect at right angles. A lens focuses the laser beam. b) Pulsed design where the laser and electron beams counter-propagate. Curved mirrors focus the laser beam and have holes to allow the electron beam to pass through. Flat mirrors and a separate lens could also be used. The part of the laser beam that passes through the hole in the bottom mirror is not depicted. c) Continuous wave design using two mirrors to form a Fabry-Perot cavity.

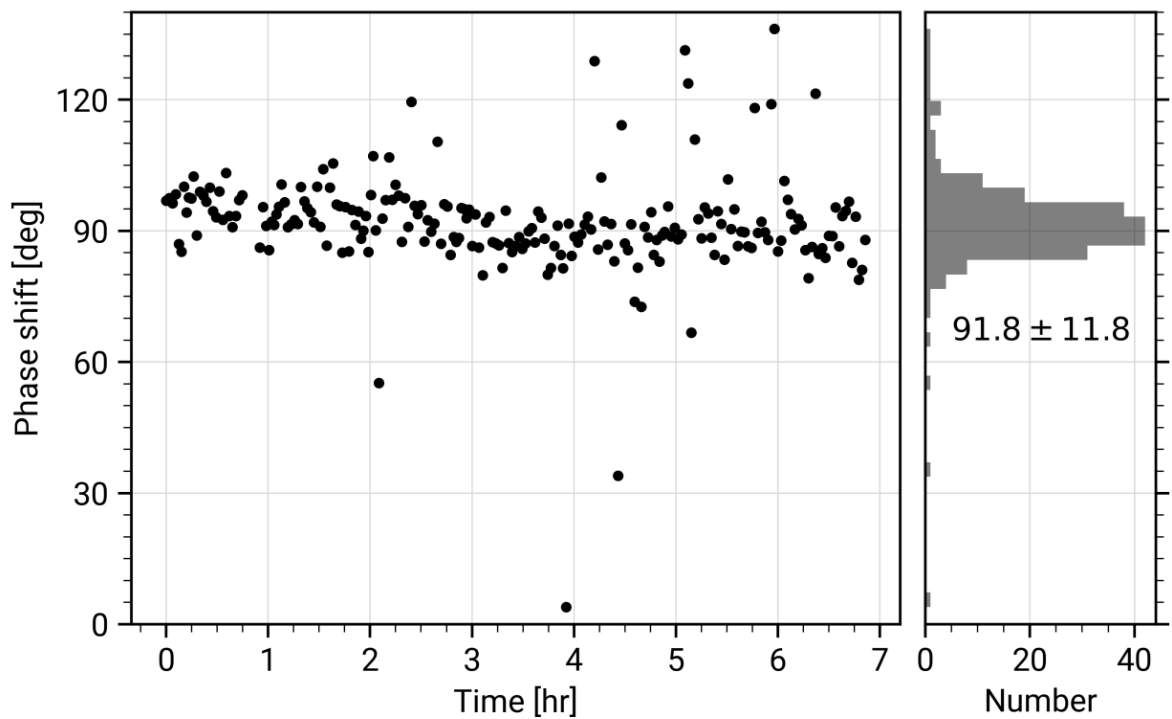

**Figure 3: Phase shift stability of a continuous-wave laser phase plate.** Fitted phase shift values of every image collected in a laser phase plate single-particle analysis dataset of apoferritin, as a function of time. The right panel shows a histogram of the values in the left panel, with a mean phase shift of 91.8° and a standard deviation of 11.8°.

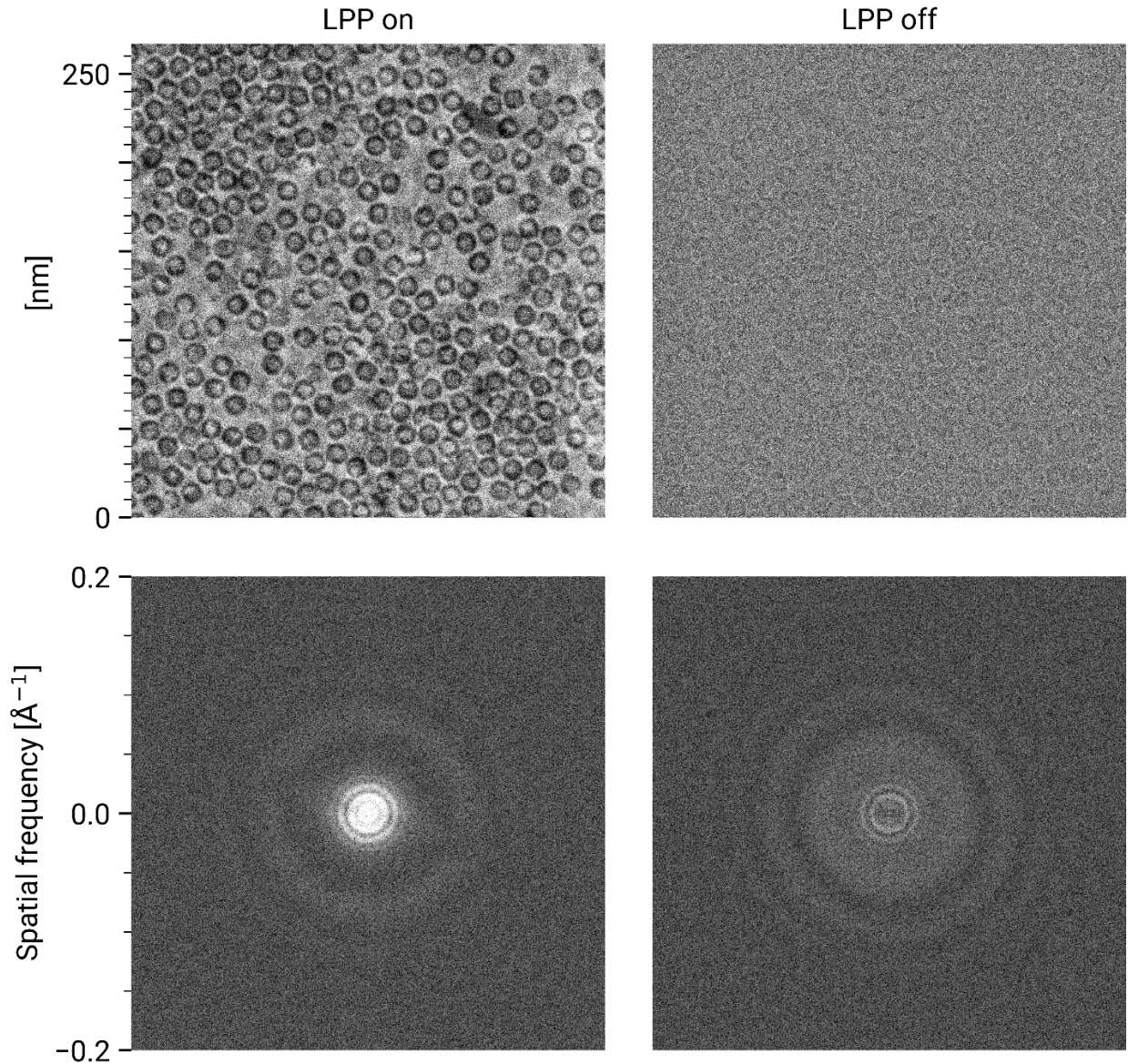

**Figure 4: Comparison between laser phase plate and defocus phase contrast.** Images (top row) and their FFTs (bottom row) of an apoferritin cryo-EM sample with the laser phase plate (LPP) on (left column) and off (right column). The defocus of the LPP on image is 638nm with a phase shift of 96°. The defocus of the "LPP off" image is 755nm. The grayscale of both images extends from 0.8-1.2, where 1 is the mean pixel value of each image. Note that both images are displayed at lower than native resolution, which somewhat reduces the apparent shot noise.